\documentclass[a4paper,11pt]{article}
\usepackage{pos}

\usepackage[symbol]{footmisc}
\usepackage[english]{babel}
\usepackage{graphicx}
\usepackage{graphics}
\usepackage{braket}
\usepackage{bbold}
\usepackage{amsmath}
\usepackage{nicefrac}
\usepackage{dcolumn}
\usepackage{bm}
\usepackage{slashed}
\usepackage{datetime}
\usepackage{mciteplus}
\usepackage{multirow}
\usepackage{bbold}
\usepackage{siunitx}
\usepackage{booktabs}
\usepackage{color, soul}
\usepackage[usenames,dvipsnames]{xcolor}
\usepackage{float}
\usepackage[utf8]{inputenc}
\usepackage[normalem]{ulem}
\usepackage{mathtools}
\usepackage{setspace}

\renewcommand{\thefootnote}{\fnsymbol{footnote}}

\newcommand{\LQCD}{\Lambda_{\rm QCD}}

\newcommand{\NLLp}{{\rm NLL/NLO^+}}

\newcommand{\NLLm}{{\rm NLL/NLO^-}}

\newcommand{\DY}{\Delta Y}

\newcommand{{\HFNRevo}}{\tt HF-NRevo}

\title{Matching NLL to NLO in Higgs and Z plus jet at the LHC and FCC}
\ShortTitle{Matching NLL to NLO in Higgs and Z plus jet at the LHC and FCC}

\author*[a]{Francesco Giovanni Celiberto}
\author[b,c]{Luigi Delle Rose}
\author[b,c]{Alessandro Papa}

\affiliation[a]{Universidad de Alcalá (UAH), Departamento de Física y Matemáticas, Campus Universitario, \\ Alcalá de Henares, E-28805, Madrid, Spain}

\affiliation[b]{Dipartimento di Fisica, Università della Calabria, Arcavacata di Rende, I-87036, Cosenza, Italy}

\affiliation[c]{
INFN, Gruppo Collegato di Cosenza, Arcavacata di Rende, I-87036, Cosenza, Italy}

\emailAdd{francesco.celiberto@uah.es}
\emailAdd{luigi.dellerose@unical.it}
\emailAdd{alessandro.papa@fis.unical.it}

\abstract{We present updated predictions for rapidity and transverse-momentum spectra in Higgs-plus-jet production at proton colliders, combining NLO fixed-order QCD with next-to-leading energy-logarithmic resummation.
Preliminary results for the $Z$-boson case are also discussed.
Our study underscores the importance of improving fixed-order predictions for Higgs- and $Z$-plus-jet observables to meet the precision demands of Higgs and electroweak measurements at the LHC and future FCC energies.}

\FullConference{The European Physical Society Conference on High Energy Physics (EPS-HEP2025)\\
7-11 July 2025\\
Marseille, France\\}

\begin{document}
\maketitle

\section{Introduction}
\label{sec:introduction}

A precise theoretical description of Higgs- and $Z$-boson production at the LHC and, in perspective, at the FCC~\cite{FCC:2025lpp,FCC:2025uan,FCC:2025jtd}, requires the \emph{all-order} resummation of logarithms enhanced at high energies.
In this work, we focus on the \emph{semi-hard} regime, characterized by a pronounced hierarchy of scales, $\sqrt{s} \gg \{\mu_i\} \gg \LQCD$, where $\sqrt{s}$ the total center-of-mass energy, $\{\mu_i\}$ denotes a set of representative hard momenta, and $\LQCD$ the QCD hadronization scale. 
This energy separation gives rise to large logarithms of energy that require resummation for accurate theoretical predictions.

The Balitsky-Fadin-Kuraev-Lipatov (BFKL) approach~\cite{Fadin:1975cb,Balitsky:1978ic} provides a framework to resum these logarithms at leading-logarithmic (LL) and next-to-leading logarithmic (NLL) orders.
Beyond resummation, the BFKL formalism enables access to the low-$x$ regime of gluon densities inside the proton, a region of particular interest for high-energy QCD phenomenology~\cite{Bacchetta:2020vty,Bacchetta:2024fci,Bacchetta:2021lvw,Bacchetta:2021twk,Bacchetta:2024uxb,Arbuzov:2020cqg,Celiberto:2021zww,Amoroso:2022eow,Bolognino:2018rhb,Bolognino:2021niq,Hentschinski:2022xnd,Celiberto:2019slj}.

A natural testing ground for this dynamics is offered by semi-inclusive processes involving the production of two final-state particles with high transverse mass, well separated in rapidity by a large interval $\DY$. These configurations are ideal to probe multi-Regge kinematics and to enhance the sensitivity to energy logarithms.
To describe such two-particle final states in a reliable and systematic way, one must adopt a \emph{multilateral} factorization strategy, able to combine collinear QCD dynamics with high-energy resummation.

To this end, the \emph{hybrid} factorization (HyF) formalism was developed~\cite{Celiberto:2020tmb,Bolognino:2021mrc}, extending earlier frameworks to enable a unified treatment of both dynamics. For single-particle production, related approaches can be found in~\cite{vanHameren:2022mtk,Bonvini:2018ixe,Silvetti:2022hyc}. In HyF, cross sections are expressed as convolutions involving a universal NLL BFKL Green’s function, which acts analogously to a Sudakov radiator in soft-gluon resummation, and two process-dependent emission functions.

Each emission function further factorizes into a convolution of collinear PDFs and hard-scattering coefficient functions evaluated with one off-shell leg.
The highest accuracy currently achievable within this setup is NLL/NLO, where both emission coefficients are computed at NLO fixed order.
Efforts to extend the high-energy reach of impact-factor calculations beyond the current NLL/NLO precision include the recent determination of the two-loop Higgs impact factor in the Regge limit within an effective-field-theory framework, at next-to-NLL accuracy~\cite{DelDuca:2025vux}.
However, in practical applications, one often employs a mixed accuracy, dubbed $\NLLm$, where one coefficient is evaluated at NLO and the other at LO, matched with the Green’s function at full NLL level.

The HyF framework has been successfully applied to a wide array of observables: Mueller-Navelet jet correlations within full NLL accuracy~\cite{Ducloue:2013hia,Celiberto:2015yba,Celiberto:2015mpa,Celiberto:2016ygs,Celiberto:2017ius,Caporale:2018qnm,Celiberto:2022gji}, forward Drell-Yan~\cite{Celiberto:2018muu,Golec-Biernat:2018kem}, light~\cite{Celiberto:2016hae,Celiberto:2017ptm,Bolognino:2018oth,Bolognino:2019yqj,Bolognino:2019cac,Celiberto:2020rxb,Celiberto:2022kxx} and heavy-light~\cite{Celiberto:2017nyx,Bolognino:2019yls,Bolognino:2019ccd,AlexanderAryshev:2022pkx,Celiberto:2021dzy,Celiberto:2021fdp,Celiberto:2022zdg,Celiberto:2022keu,Celiberto:2024omj,Anchordoqui:2021ghd,Feng:2022inv,Celiberto:2022grc} hadrons, quarkonia~\cite{Boussarie:2017oae,Chapon:2020heu,Celiberto:2022dyf,Celiberto:2023fzz,Celiberto:2022grc}, rare~\cite{Celiberto:2025ogy} and exotic~\cite{Celiberto:2023rzw,Celiberto:2024mab,Celiberto:2024mrq,Celiberto:2024beg,Celiberto:2025dfe,Celiberto:2025ziy} hadrons, among others.

In the present study, we explore the application of this formalism to Higgs-plus-jet production in proton collisions~\cite{Celiberto:2020tmb,Celiberto:2023rtu}, a channel that has been extensively studied in perturbative QCD at NNLO~\cite{Chen:2014gva,Boughezal:2015dra,Dawson:2022zbb}, and via next-to-NLL resummation of transverse-momentum logarithms~\cite{Monni:2019yyr}. These existing approaches, however, do not include the resummation of logarithms enhanced at high energies or account for gluon radiation across large rapidity intervals.

To bridge this gap, we introduce the POWHEG+JETHAD method, a novel matching strategy that combines NLO fixed-order matrix elements, generated through the POWHEG formalism, with the NLL resummation of energy logarithms encoded in HyF.
This approach enables the generation of realistic, resummation-improved predictions for observables sensitive to large rapidity gaps and semi-hard kinematics, and provides a promising pathway toward precision Higgs and electroweak physics at both the LHC and the FCC.

\section{Towards Higgs-plus-jet production at NLL/NLO}
\label{sec:matching}

\begin{figure*}[!t]
\centering

\includegraphics[scale=0.37,clip]{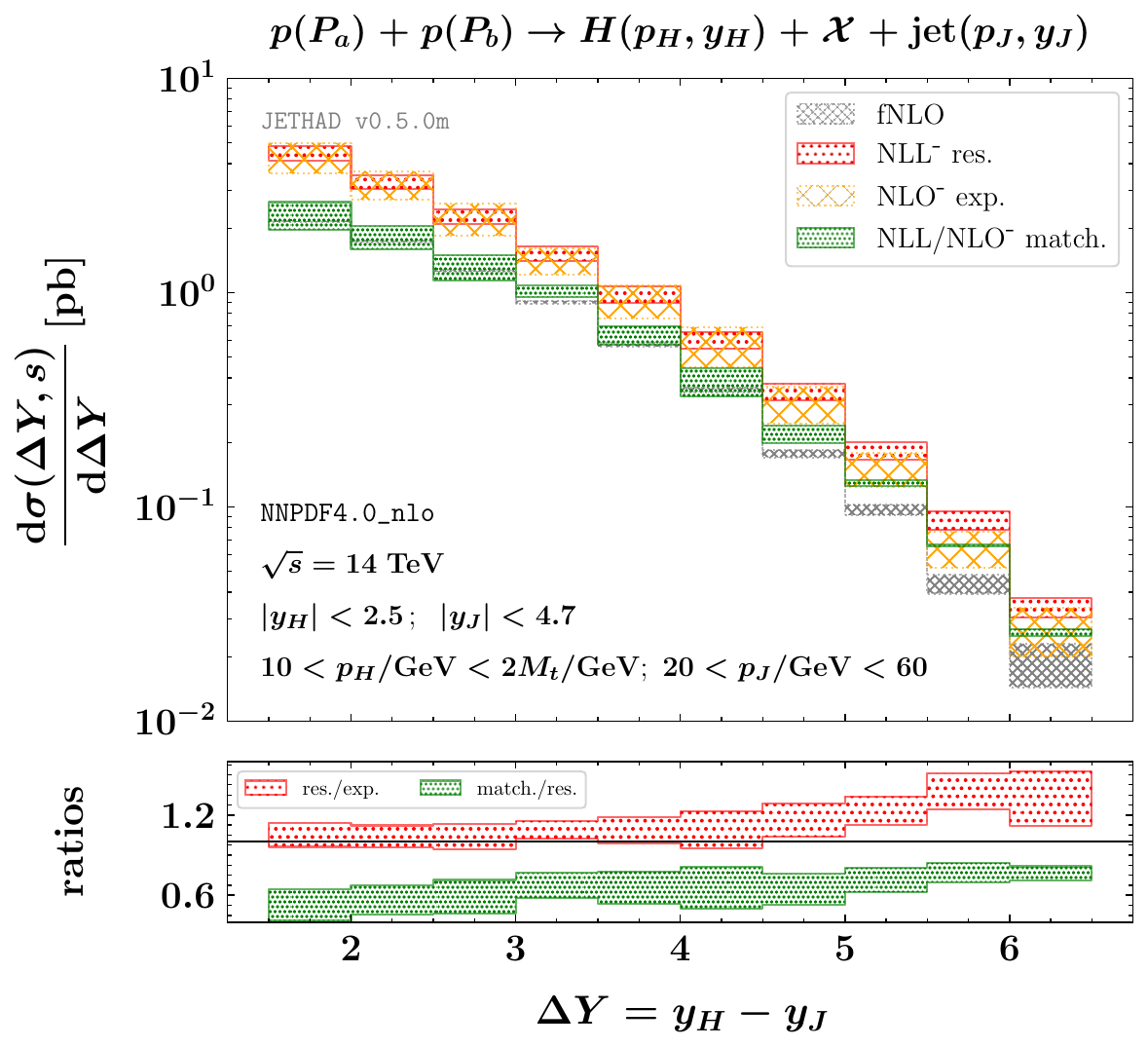}
 \hspace{0.00cm}
\includegraphics[scale=0.37,clip]{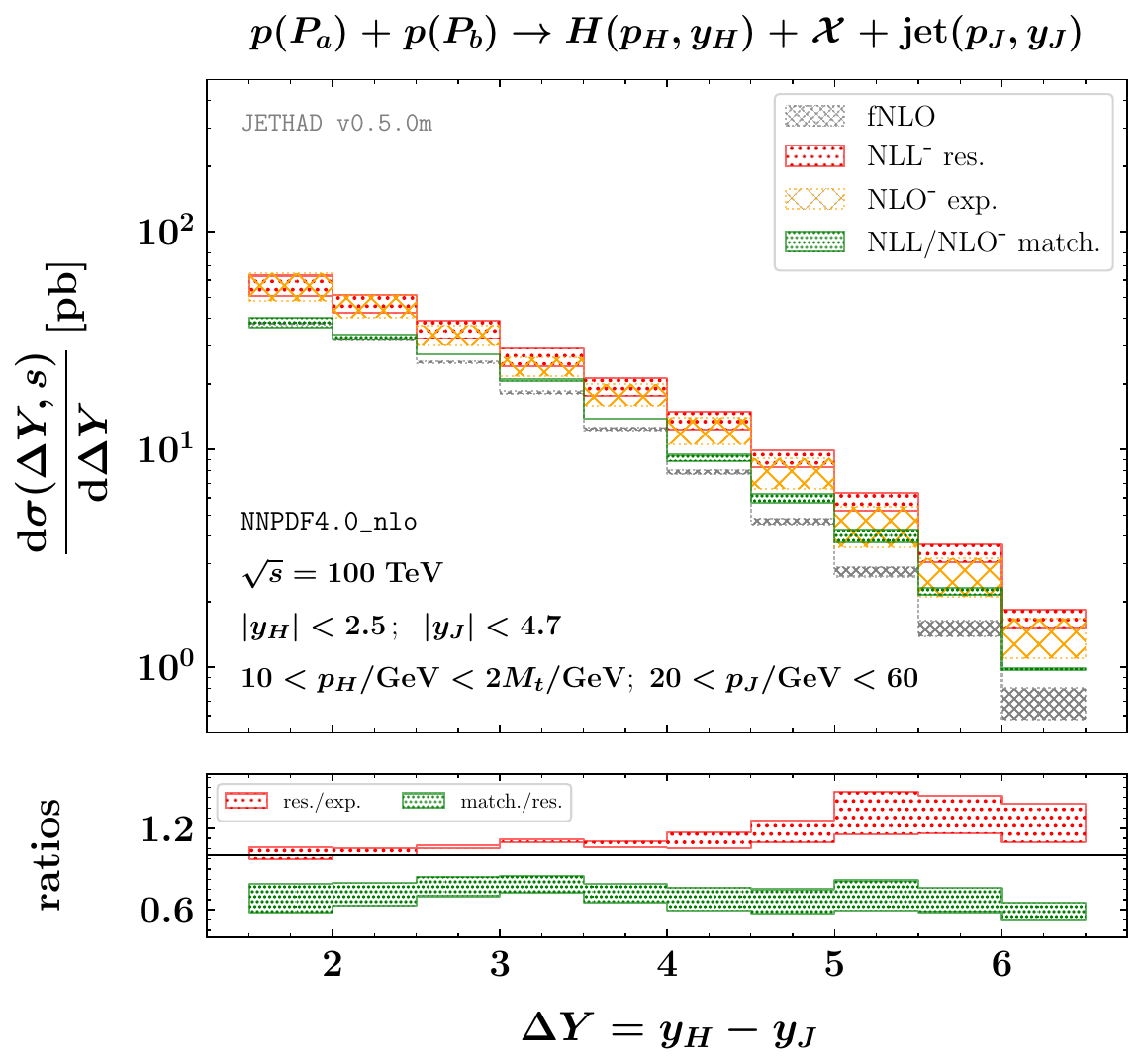}

\caption{Higgs-plus-jet $\DY$ spectrum at $14$ TeV LHC (left) and $100$ TeV nominal FCC (right) energies.
Uncertainty bands show $\mu_{R,F}$ variation in the $1 < C_{\mu} < 2$ range. 
Text boxes refer to kinematic cuts.
}

\label{fig:I}
\end{figure*}

\begin{figure*}[!t]
\centering

\includegraphics [scale=0.36,clip]{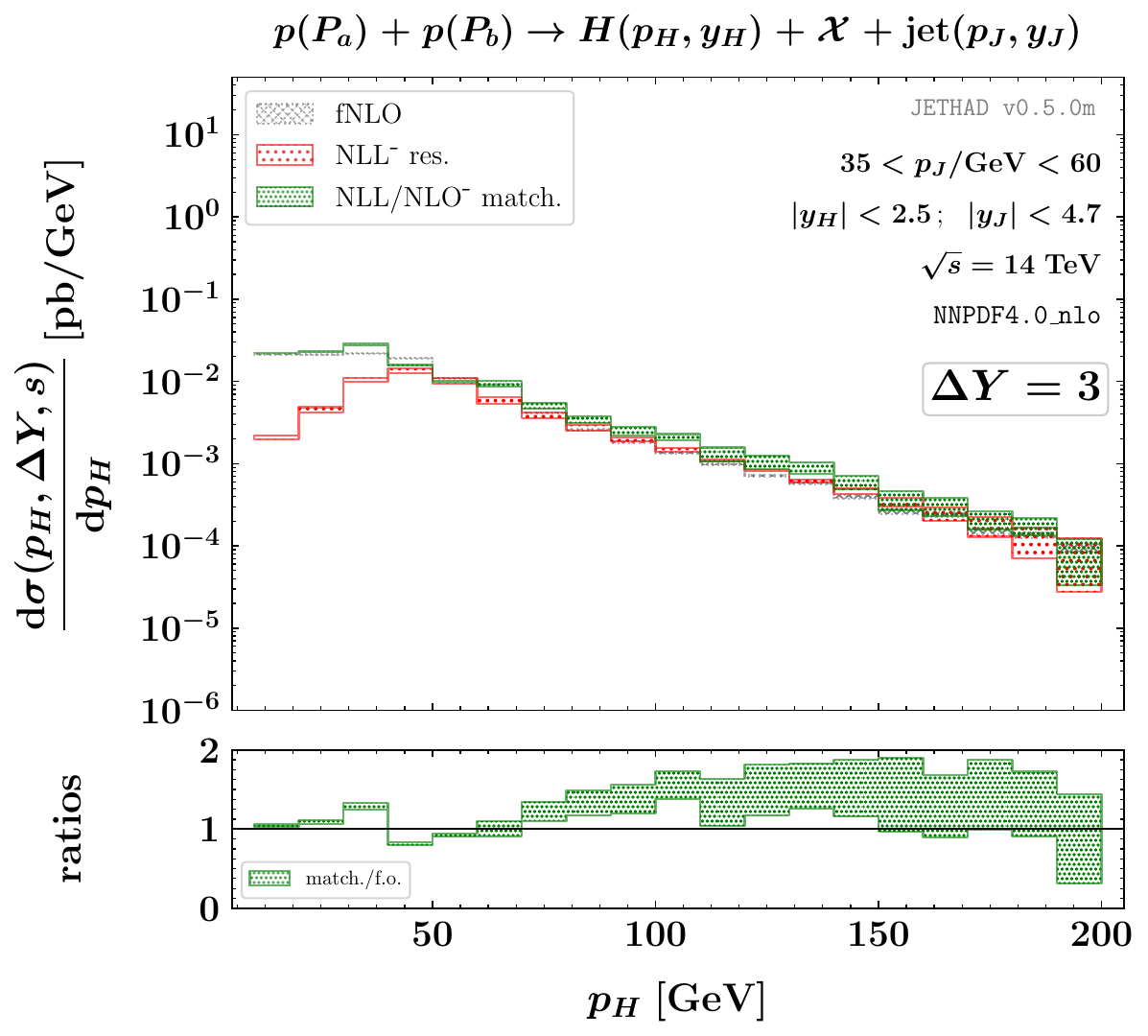}
 \hspace{0.00cm}
\includegraphics[scale=0.36,clip]{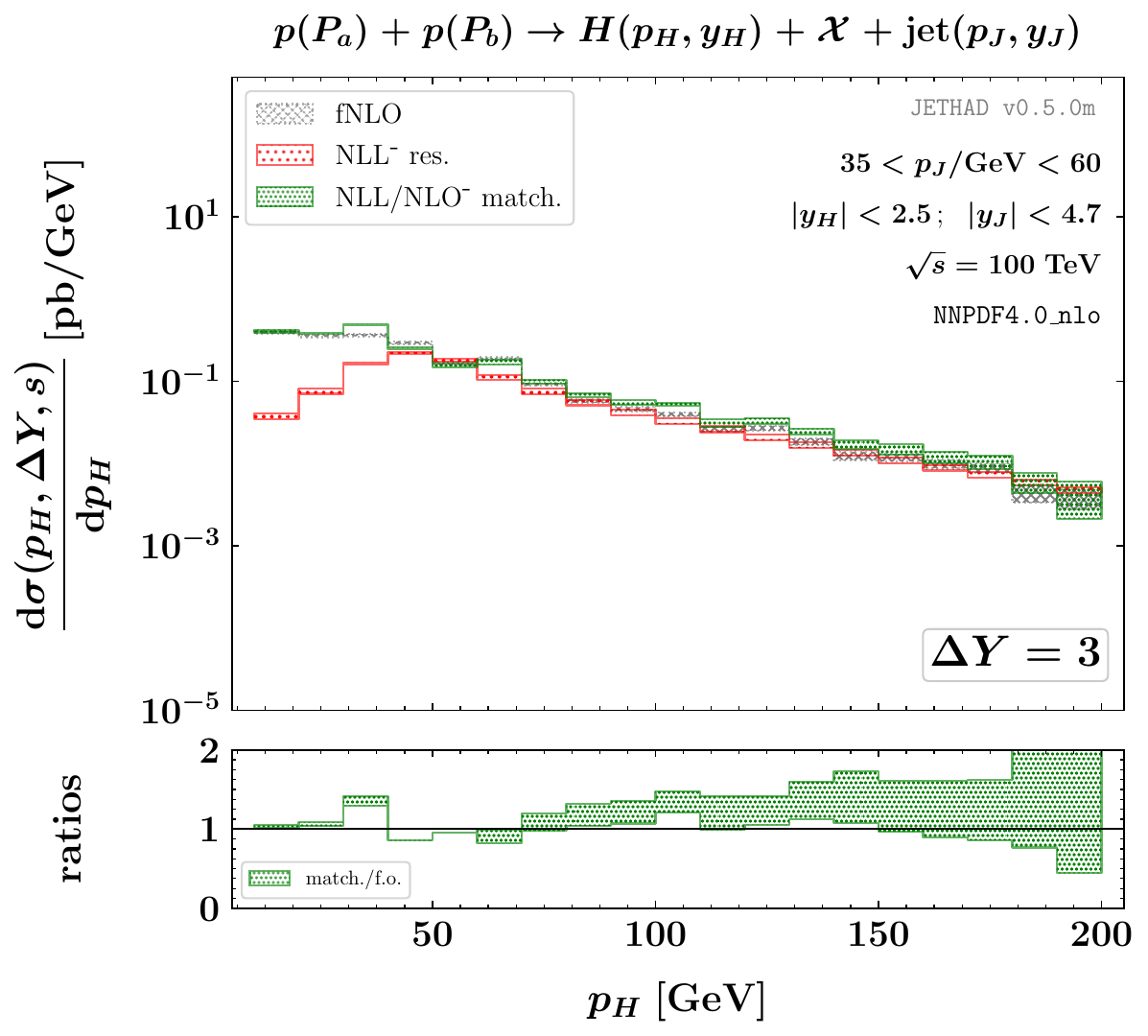}

\caption{Higgs-plus-jet $p_H$ spectrum at $14$ TeV LHC (left) and $100$ TeV nominal FCC (right) energies.
Uncertainty bands show $\mu_{R,F}$ variation in the $1 < C_{\mu} < 2$ range. 
Text boxes refer to kinematic cuts.
}

\label{fig:pT}
\end{figure*}

Preliminary investigations of the Higgs transverse-momentum ($p_H$) distribution in the inclusive Higgs-plus-jet channel at both LHC~\cite{Celiberto:2020tmb} and FCC~\cite{Celiberto:2023rtu} energies, performed within the HyF framework, have demonstrated notable stability under radiative corrections and scale variations.
These results provide a solid benchmark for the application of high-energy resummation techniques to semi-hard observables involving Higgs bosons.
However, despite this theoretical robustness, a significant mismatch was observed between HyF predictions and the corresponding pure fixed-order calculations, particularly in the intermediate-$p_H$ region where both approaches should, in principle, be simultaneously valid.

To address this discrepancy and improve the theoretical consistency of the predictions, we have developed a dedicated \emph{matching} strategy that combines NLO fixed-order calculations with next-to-leading logarithmic NLL high-energy resummation.
This approach is based on the exact subtraction of the \emph{double-counted} contributions present at the $\NLLm$ level of accuracy, ensuring that only genuinely resummed logarithmic terms are retained on top of the fixed-order prediction.
Such a subtraction scheme is essential to avoid overlapping between the two components and to preserve the correct normalization and shape of the differential distributions.

At present, the full NLO expression for the Higgs emission function~\cite{Hentschinski:2020tbi,Celiberto:2022fgx,Nefedov:2019mrg,Celiberto:2024bfu} has not yet been incorporated into the JETHAD framework~\cite{Celiberto:2020wpk,Celiberto:2022rfj,Celiberto:2023fzz,Celiberto:2024mrq,Celiberto:2024swu}.
As a result, we work within a $\NLLm$ setup, where one of the emission functions is evaluated at NLO and the other one remains at LO accuracy.
This configuration already captures the dominant high-energy dynamics and allows for meaningful phenomenological applications.

The structure of our matching procedure and its implementation within the POWHEG+JETHAD environment are detailed in Refs.~\cite{Celiberto:2023uuk_article,Celiberto:2023eba_article,Celiberto:2023nym,Celiberto:2024mdt}, and will serve as the foundation for upcoming developments toward full $\NLLp$ predictions.
For clarity and compactness, the structure of our matching strategy can be summarized by the following (somewhat schematic) formula, which captures the essential ingredients of our implementation

\begin{equation}
\label{eq:matching}
\begin{split}
 \hspace{-0.255cm}
 \underbrace{{\rm d}\sigma^{{{\rm NLL/NLO}}^{\boldsymbol{-}}}(\Delta Y, \varphi, s)}_{\text{\colorbox{OliveGreen}{\textbf{\textcolor{white}{NLL/NLO$^{\boldsymbol{-}}$}}} {\tt POWHEG+JETHAD}}} 
 = 
 \underbrace{{\rm d}\sigma^{\rm NLO}(\Delta Y, \varphi, s)}_{\text{\colorbox{gray}{\textcolor{white}{\textbf{NLO}}} {\tt POWHEG} w/o PS}}
 +\; 
 \underbrace{\underbrace{{\rm d}\sigma^{{{\rm NLL}}^{\boldsymbol{-}}}(\Delta Y, \varphi, s)}_{\text{\colorbox{red}{\textbf{\textcolor{white}{NLL$^{\boldsymbol{-}}$ resum}}} (HyF)}}
 \;-\; 
 \underbrace{\Delta{\rm d}\sigma^{{{\rm NLL/NLO}}^{\boldsymbol{-}}}(\Delta Y, \varphi, s)}_{\text{\colorbox{orange}{\textbf{NLL$^{\boldsymbol{-}}$ expanded}} at NLO}}}_{\text{\colorbox{NavyBlue}{\textbf{\textcolor{white}{NLL$^{\boldsymbol{-}}$}}} {\tt JETHAD} w/o NLO$^{\boldsymbol{-}}$ double counting}}
 \,.
\end{split}
\end{equation}

A matched observable, ${\rm d}\sigma^{{{\rm NLL/NLO}}^{\boldsymbol{-}}}$, computed at $\NLLm$ via the POWHEG+JETHAD framework, is expressed as the sum of two components.
The first is the fixed-order NLO result (gray), generated by POWHEG~\cite{Hamilton:2012rf,Bagnaschi:2023rbx,Banfi:2023mhz}, and shown without the application of a \emph{parton shower} (PS)~\cite{Alioli:2022dkj,vanBeekveld:2022zhl,FerrarioRavasio:2023kyg}.
The second is the high-energy resummed correction (blue), obtained from JETHAD, and defined as the difference between the $\rm NLL^-$ HyF resummed contribution (red) and its NLO expansion (orange), thus ensuring the consistent removal of any \emph{double counting}.

Building on the preliminary results presented in Refs.~\cite{Celiberto:2023uuk_article,Celiberto:2023eba_article,Celiberto:2023nym}, we extend the analysis to 100~TeV nominal FCC energies.
In particular, we present predictions for the rapidity separation spectrum, $\DY$, shown in Fig.~\ref{fig:I}, and for the Higgs transverse momentum distribution, $p_H$, at fixed $\DY = 3$, displayed in Fig.~\ref{fig:pT}.
Left panels are for results at 14~TeV LHC, whereas right panels are for results at 100~TeV (nominal) FCC.

In Figs.~\ref{fig:I} and~\ref{fig:pT}, the labels used in the main panels and the ratio plots follow the structure of Eq.~\eqref{eq:matching}. 
The ``fNLO'' (gray) curve represents the fixed-order prediction obtained via POWHEG without the inclusion of any PS effect, and it corresponds to the term ${\rm d}\sigma^{\rm NLO}$ in Eq.~\eqref{eq:matching}. 
The ``NLL$^-$~res'' (red) curve denotes the full high-energy resummed prediction from JETHAD, representing the ${\rm d}\sigma^{\rm NLL^-}$ component. 
Its NLO expansion, shown as ``NLL$^-$~exp'' (orange), is the $\mathcal{O}(\alpha_s^3)$ truncation of the NLL$^-$ series and defines the subtraction term $\Delta{\rm d}\sigma^{\rm NLL/NLO^-}$.
Finally, the ``NLL/NLO$^-$~match'' (green) curve corresponds to the complete matched observable ${\rm d}\sigma^{\rm NLL/NLO^-}$, ensuring the removal of double counting between fixed-order and resummed contributions.

The ratio plots beneath each panel further clarify the interplay between the various components.
The ``res./exp.'' ratio in Fig.~\ref{fig:I} quantifies the genuine resummation enhancement by comparing the full resummed prediction to its truncated expansion, while the ``match./res.'' assesses the weight of the matching with respect to the purely resummed calculation.
In Fig.~\ref{fig:pT}, the ``match./f.o.'' ratio highlights the net effect of high-energy resummation over the fixed-order result.
This explicit mapping between the graphical labels and the theoretical expression in Eq.~\eqref{eq:matching} facilitates a transparent interpretation of the phenomenological results.

The robustness and consistency of our matching strategy are clearly reflected in the numerical results shown in the figures.
Let us first consider the behavior of the rapidity-separation spectrum, $\DY$ (Fig.~\ref{fig:I}).
By analyzing the ratios displayed in the ancillary plots beneath the main distributions, we observe a monotonic enhancement of the NLL$^-$ resummed predictions over their NLO$^-$ expanded counterparts as $\DY$ increases.
This trend confirms that the impact of high-energy resummation becomes progressively more relevant in the large-rapidity regime, as expected from BFKL dynamics.
The presence of increasingly large logarithms of energy in this kinematic region amplifies the resummation effects, and their proper treatment is essential to obtain physically meaningful predictions.

Moreover, the ratio between the matched $\NLLm$ results and the pure NLL$^-$ resummed ones gradually approaches unity at large $\DY$, indicating that, in this limit, the contribution from the fixed-order baseline becomes less dominant compared to the resummed enhancement.
This convergence further substantiates the validity of our matching procedure and highlights the growing importance of high-energy logarithms in semi-hard observables.
The effect is particularly visible at $\sqrt{s} = 14$~TeV, where the transition between fixed-order and resummed behavior is smoother and better resolved.
At $\sqrt{s} = 100$~TeV, the same pattern is qualitatively present but less pronounced, due to the broader phase space available and the relative suppression of the resummation effect over a wider kinematic range.

Turning to the transverse-momentum spectrum of the Higgs boson, $p_H$ (Fig.~\ref{fig:pT}), we find that high-energy resummation leads to a substantial enhancement over the fixed-order result in the region around the spectral peak.
In particular, the matched-to-fixed-order ratio indicates corrections ranging from 30 to 50\% in the peak and near-peak regions.
This is consistent with earlier analyses~\cite{Celiberto:2020tmb}, which pointed out that the $p_H$ distribution, especially when the Higgs and the accompanying jet carry similar transverse momenta, is sensitive to logarithms of the type $\ln(\hat s / p_H^2)$, which are precisely those resummed by the BFKL approach.

As we move toward higher $p_H$ values, the uncertainty bands of our matched predictions widen significantly.
This broadening reflects the growing influence of additional classes of logarithms not captured by the pure energy resummation.
In particular, large collinear (DGLAP-type) logarithms and near-threshold effects~\cite{Bonciani:2003nt,deFlorian:2005fzc,Muselli:2017bad} become increasingly important at high transverse momentum.
These effects slow down the convergence of the high-energy series and signal the need for complementary resummation techniques in that region.

Nonetheless, a key result of our analysis is that, unlike previous predictions obtained purely within the HyF framework, notably those shown in Fig.~8 of Ref.~\cite{Celiberto:2020tmb}, our $\NLLm$ matching procedure leads to a visible reduction of the tension between the resummed and fixed-order results in the high-$p_H$ tail.
This improved agreement confirms that the POWHEG+JETHAD matching not only preserves the theoretical consistency of the underlying factorization schemes but also delivers more precise and phenomenologically reliable predictions for observables relevant to Higgs physics in the semi-hard domain.

\section{Conclusions and outlook}
\label{sec:conclusions}

We have introduced a novel matching strategy that combines the POWHEG formalism~\cite{Hamilton:2012rf,Bagnaschi:2023rbx,Banfi:2023mhz} with the JETHAD framework~\cite{Celiberto:2020wpk,Celiberto:2022rfj,Celiberto:2023fzz,Celiberto:2024mrq,Celiberto:2024swu}, with the goal of consistently merging NLO fixed-order predictions with the resummation of high-energy logarithms at NLL accuracy and beyond, reaching the so-called NLL/NLO$^+$ precision level.
This approach provides a fully differential and resummation-improved description of semi-hard observables, and sets the stage for systematically improving theoretical predictions in forward and large-rapidity-gap processes.

Looking ahead, several directions are being pursued to increase the precision and generality of the framework. 
These include the numerical implementation of full NLO corrections to the Higgs emission function, building on existing studies~\cite{Hentschinski:2020tbi,Celiberto:2022fgx,Nefedov:2019mrg,Celiberto:2024bfu}, as well as the inclusion of finite heavy-quark mass effects in the matrix elements, following recent developments in massive-loop calculations~\cite{Jones:2018hbb,Bonciani:2022jmb,Celiberto:2024bbv,Celiberto:2025ece}.
In parallel, work is ongoing to generalize the current setup to neutral electroweak boson production, in particular to extend the formalism to the $Z$-boson channel, which is experimentally accessible and theoretically complementary to the Higgs case.



\begingroup
\setstretch{0.6}
\bibliographystyle{bibstyle}
\bibliography{biblography}
\endgroup

\end{document}